\documentclass[runningheads]{llncs}
\usepackage[T1]{fontenc}
\usepackage{graphicx}
\usepackage{amsmath}
\usepackage{amssymb}
\usepackage{etoolbox}
\usepackage{siunitx}
\usepackage{booktabs}

\usepackage{times}
\usepackage[misc]{ifsym}
\usepackage{epsfig}
\usepackage{subfigure}
\usepackage{multicol,multirow}
\usepackage{cite}
\usepackage{bigstrut}
\usepackage{floatrow}
\usepackage{rotating}
\usepackage{adjustbox}

\begin{document}
\title{Efficient Monaural Speech Enhancement using Spectrum Attention Fusion}
%
%

\author{Jinyu Long\textsuperscript{(\,\Letter\,)}\inst{1} \and
Jetic G\=u\inst{2} \and
Binhao Bai\inst{1} \and
Zhibo Yang\inst{1} \and
Ping Wei\inst{4} \and
Junli Li\textsuperscript{(\,\Letter\,)}\inst{1,3}}

\institute{
School of Computing Science, Sichuan Normal University, Chengdu 610066, China 
\email{sicnu\_long@stu.sicnu.edu.cn, lijunli@sicnu.edu.cn}\\
\and School of Computing Science, Simon Fraser University, Burnaby, BC V5A 1S6, Canada 
\and Visual Computing and Virtual Reality Key Laboratory of Sichuan, Sichuan Normal University, Chengdu 610068, China 
\and Sichuan Key Laboratory of Translational Medicine of Traditional Chinese Medicine, Chengdu 610041, China }
\authorrunning{J. Long et al.}
%
\maketitle              
\begin{abstract}
Speech enhancement is a demanding task in automated speech processing pipelines, focusing on separating clean speech from noisy channels. Transformer based models have recently bested RNN and CNN models in speech enhancement, however at the same time they are much more computationally expensive and require much more high quality training data, which is always hard to come by.
In this paper, we present an improvement for speech enhancement models that maintains the expressiveness of self-attention while significantly reducing model complexity, which we have termed Spectrum Attention Fusion.
We carefully construct a convolutional module to replace several self-attention layers in a speech Transformer, allowing the model to more efficiently fuse spectral features.
Our proposed model is able to achieve comparable or better results against SOTA models but with significantly smaller parameters (0.58M) on the Voice Bank + DEMAND dataset. 

\end{abstract}
\section{Introduction}
Real-world acoustic environments are often contaminated with different forms of background disturbances, which can significantly hinder speech-based applications such as phone calls and Automatic Speech Recognition (ASR).
Monaural speech enhancement techniques aim to recover clean speech from the original noisy input, by often isolating speech related information from the noisy background.
While conventional statistical methods are primarily based on specific speech and noise characteristics \cite{untwale2015survey}, in recent years neural network based models have been shown to be much more versatile at handling non-stationary noise as well as more challenging scenarios such as with low Signal-to-Noise Ratio (SNR) and reverberation\cite{wang2018supervised}.

In a neural speech processing model, a common approach is to directly encode a time-frequency representation (i.e. spectrogram, Mel spectrogram) as input\cite{purwins2019deep}.After the transformation, the spectrogram, either in single vector forms or as 2D image slices, is fed directly into the encoding layer of the network.
To extract features from diverse spectra, previous studies have used Convolutional Neural Networks (CNNs) to capture local information in neighbouring sub-bands in a duration \cite{park2016fully, kounovsky2017single}, Recurrent Neural Networks (RNNs) for modelling sequential features in frequency or time dimensions \cite{hao2021fullsubnet, valin2018hybrid, gao2018densely}, and hybrid architectures to collaboratively boost capacity \cite{tan2018convolutional, strake2020fully}.
More recently, Transformer models built more on dynamic global context interactions \cite{vaswani2017attention} have shown even more remarkable results in extracting features from spectra \cite{baevski2020wav2vec, yu2022setransformer}.

Despite being highly expressive, a major problem with transformers is their quadratic computational complexity, which can be a constraint when dealing with long inputs.
Recently proposed Computer Vision model Conv2Former \cite{hou2022conv2former} attempts to address this issue by computing the Hadamard product between a large-kernel convolution's and multi-head values to mimic self-attention, dramatically reducing parameter size and computation cost.
This shows that vision transformers can indeed be optimized to work more efficiently, and we hypothesise that this may be the same in the speech domain.

In this work, we introduce a new approach named Spectrum Attention Fusion, which refines and enhances the Convolutional Modulation from Conv2Former to more effectively integrate spectral features in speech processing.
Our Spectrum Attention Fusion module augments the speech Transformer's self-attention mechanism\cite{yu2022setransformer} with convolutional layers, so it can more efficiently fuse these features while retaining access to a more global context.
With significantly fewer parameters (0.58M) than a full Transformer, our proposed model achieves better performance than a vanilla Transformer, and remains competitive or better (2.84 WB-PESQ, 94.3\% STOI) on the Voice Bank + DEMAND dataset compared to other previous models including ones with much more parameters.

\section{Related Work}
Supervised speech enhancement models can generally be divided into two categories based on the kind of input the model receives after preprocessing: time-domain and frequency-domain.
Time-domain methods \cite{rethage2018wavenet, pandey2019tcnn} directly encode the normalised noisy speech waveform into the model, while frequency-domain methods \cite{park2016fully, valin2018hybrid, gao2018densely} first perform short-time Fourier transformation (STFT), then feed the spectrogram into the model.
For the latter approach, the learning objective is often to generate a noise mask \cite{wang2005ideal, sun2017multiple, williamson2015complex} that is applied on the noisy spectrogram prior to reconstruction.

The predicted noise mask is typically calculated for and applied on the magnitude spectrum, and during reconstruction needs to be combined with the phase spectrum, which is usually completely unaltered.
However, the importance of recovering phase information in perception should not be overlooked, especially when the SNR is low \cite{choi2019phase}.
One way is to utilise complex neural networks to directly process complex values, allowing them to learn features directly from both spectra \cite{hu2020dccrn, choi2019phase}, however complex neural network research is itself a challenging research direction.
Another powerful phase-aware technique is to have sub-networks for refining spectrogram \cite{li2022glance, li2021two}, where magnitude and phase information are processed with interconnections.
In this work, our modelling is based on the latter approach.

Many different types of noise typically exhibit different levels of non-stationary characteristics.
To improve a model's ability to discern noise signals and effectively capture the local spectral context, prior work has suggested the incorporation of sub-band units for extracting features from adjacent frequencies\cite{hao2021fullsubnet}, but its performance lag behind that of transformers which allows for more dynamic interactions for more complicated types of non-stationary patterns spreading across multiple frequency bands \cite{yu2022setransformer}.
Transformers as mentioned above, suffers however from high computational cost and rely on more high quality training data to achieve peak performance.
To solve this problem, we augment the self-attention mechanism to focus on specific frequency ranges while replacing more complicated heads with Temporal Convolution Network (TCN) blocks \cite{bai2018empirical}.
As shown in our experimental results, our proposed model remain highly competitive if not better than compared previous SOTA models, and the  reduced computational and memory overhead also make our model suitable for a wider range of applications, including those with limited resources or stringent efficiency constraints.

\section{METHOD}

\begin{figure*}
    \centering
    \centerline{\includegraphics[width=1\columnwidth]{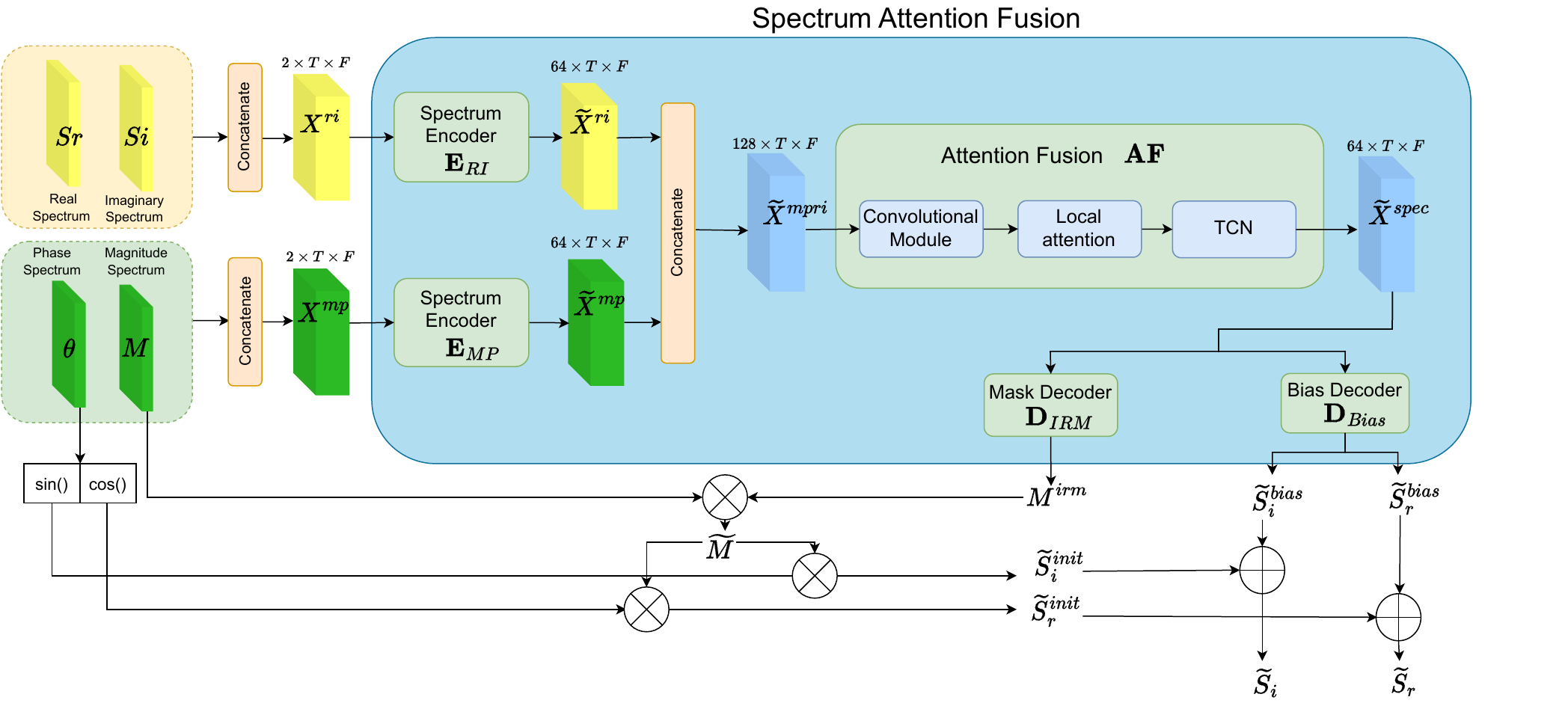}}
    \caption{
        \label{fig:overview}
        The overview of the proposed Spectrum Attention Fusion}
    \vspace{-0.4cm}
\end{figure*}

Fig.~\ref{fig:overview} presents an overview of our proposed model, which consists of three main components: spectrum encoders $\mathbf{E}_{MP}$ and $\mathbf{E}_{RI}$ for magnitude-phase and real-imaginary spectra, Spectrum Attention Fusion $\mathbf{AF}$, and decoders $\mathbf{D}_{IRM}$ and $\mathbf{D}_{Bias}$ for decoding the ideal ratio mask and bias of the estimated spectrogram, respectively.

Specifically, the proposed model accepts the magnitude, phase, real, and imaginary spectra ${M}$, ${\theta}$, ${S}_{r}$, ${S}_{i} \in \mathbb{R}^{T \times F}$ as input, where $T$ and $F$ denote the total number of frequency frames and bins, respectively.
Within the proposed $\mathbf{AF}$, the encoded representations are first fed into the Convolutional Module, then through Local Attention to perform sub-band contextual feature extraction, and finally TCN blocks for temporal feature extraction, ultimately resulting in $\widetilde{{X}}^{spec} \in \mathbb{R}^{64 \times T \times F}$.

In the decoding stage, the mask decoder $\mathbf{D}_{IRM}$ uses $\widetilde{{X}}^{spec}$ to generate the ideal ratio mask $ {M}^{irm}$, which is paired with the noisy phase $\theta$ to produce the initial estimated spectrogram $\left\{\widetilde{{S}}^{init}_r, \widetilde{{ S}}^{init}_i \right\}$, while Bias Decoder $\mathbf{D}_{Bias}$ leverages $\widetilde{{X}}^{spec}$ to compute the bias values $\left\{\widetilde{{S}}^{bias}_r, \widetilde{{S}}^{bias}_i \right\}$ of the initial estimated spectrogram. We then obtain the enhanced spectrogram $\left\{\widetilde{\mathbf{S}}_r, \widetilde{\mathbf{ S}}_i \right\}$ by refining the initial estimated spectrum with the bias values. The entire decoding process can be formulated as:

\begin{align}
    \label{eqn:1}
    \widetilde{M} = {M} \otimes {M}^{irm}\\
    \widetilde{S}^{init}_{r} = \widetilde{M} \otimes \cos\left( \theta \right)\\
    \widetilde{S}^{init}_{i} = \widetilde{M} \otimes \sin\left( \theta \right)\\
    \widetilde{S}_{r} = \widetilde{S}^{init}_{r} + \widetilde{S}^{bias}_{r}\\
    \widetilde{S}_{i} = \widetilde{S}^{init}_{i} + \widetilde{S}^{bias}_{i}
\end{align}
where the $\otimes$ denotes element-wise product.

\subsection{Spectrum Encoder}

\begin{figure}[h]
    \centering
    \centerline{\includegraphics[width=0.87\columnwidth]{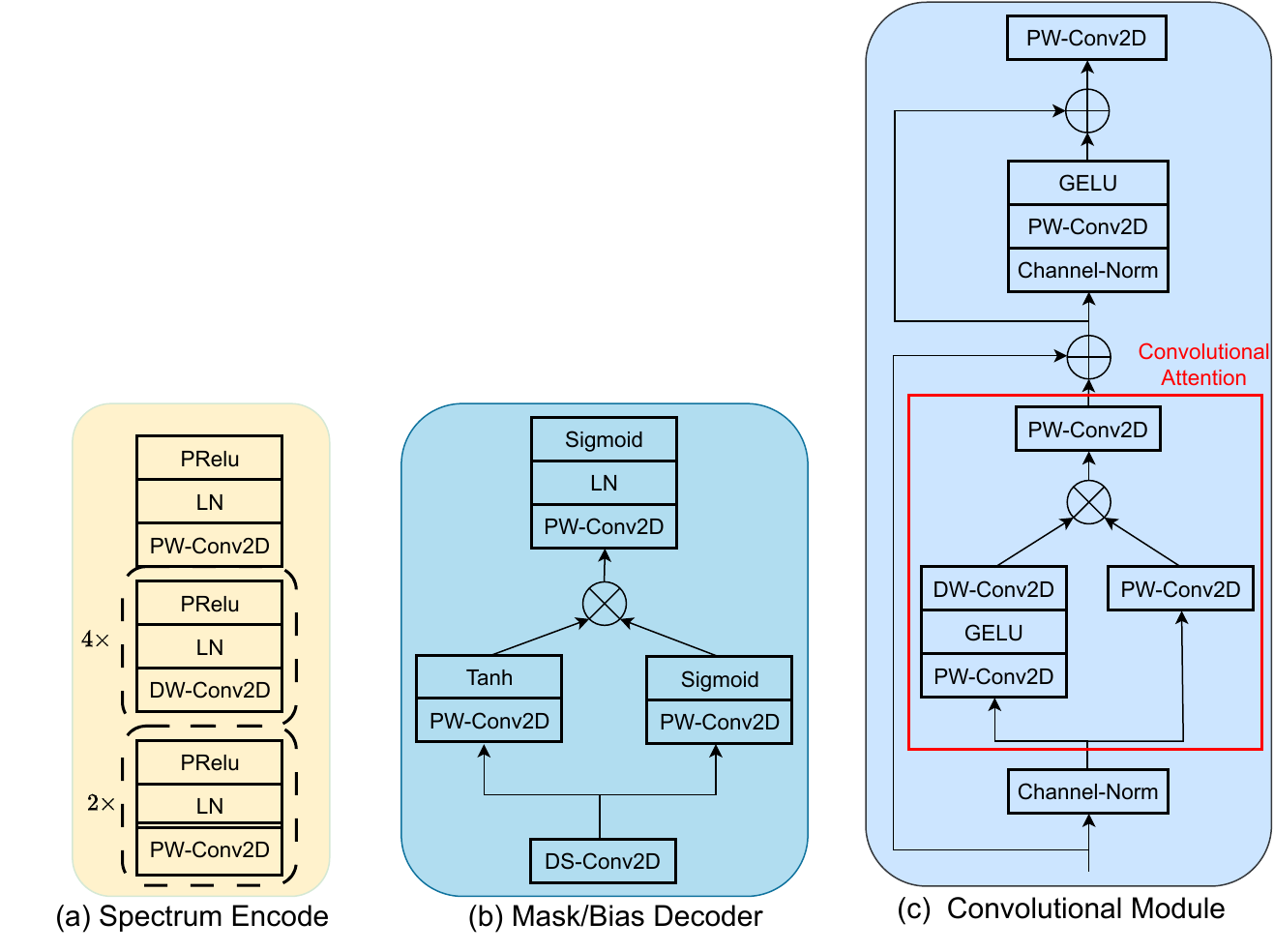}}
    \vspace{0.2cm}
    \caption{
        \label{fig:encoder}
        (a) The illustration of Spectrum Encoder. 
        (b) The illustration of the Mask/Bias Decoder.  \ \ 
        (c) The illustration of Convolutional Module. The ``DW-Conv'' indicates a depth-wise convolution. The ``PW-Conv'' indicates a point-wise convolution.The ``DS-Conv'' indicates a depth separable convolution.} 
    \vspace{-0.1cm}
\end{figure}

The encoder directly takes input representations ${M}$, ${\theta}$, ${S}_{r}$, ${S}_{i} \in \mathbb{R}^{T \times F}$ and encodes them into hidden representations for each frame.
Specifically, ${M}$, ${\theta}$ are concatenated into ${{X}}^{mp} \in \mathbb{R}^{2 \times T \times F}$; similarly, ${S}_{r}$, ${S}_{i}$ produce ${{X}}^{ri} \in \mathbb{R}^{2 \times T \times F}$, the concatenation creates a new channel dimension. By up-sampling channels of ${X}^{ri}, {X}^{mp}$ in $\mathbf{E}_{MP}$ and $\mathbf{E}_{RI}$, we can obtain $\widetilde{X}^{mp}, \widetilde{X}^{ri} \in \mathbb{R}^{64 \times T \times F}$, respectively.
Next, we concatenate ${\widetilde{X}}^{ri}$, ${\widetilde{X}}^{mp}$ to form ${\widetilde{X}}^{mpri} \in \mathbb{R}^{128 \times T \times F}$, which serves as input to Attention Fusion $\mathbf{AF}$.

As shown in Fig.~\ref{fig:encoder} (a), the Spectrum Encoder comprises two layers with 2-D point-wise convolution, layer normalisation (LN), and parametric ReLU (PReLU) activation. Subsequently, a four-layer depth-wise 2-D convolution with kernel size (1,3) is applied, along with LN and PReLU. A final 2-D point-wise convolutional layer is used, followed by LN and PReLU. This depth-separable convolution-like architecture aims to up-sample channel of spectrum features while reducing the number of parameters and computation needed.

\subsection{Attention Fusion}

As shown in Fig.~\ref{fig:encoder} (c), the Convolutional Module employs a convolutional attention mechanism to selectively concentrate on features across spectral channels.
For an input $X \in \mathbb{R}^{C \times T \times F}$, we compute the value $V$ using a point-wise convolution (i.e. linear  projection in channel dimension).
In parallel, $X$ is transformed through the same channel projection and the GELU activation function to generate the similarity score $A$ from a depth-wise convolution with an 11x11 kernel size.
The final value $Z$ is produced by taking the element-wise product of the value $V$ and the similarity score $A$, followed by another channel projection.
Convolutional attention can be formulated as follows:

\begin{align}
    \label{eqn:conv}
    V &= \mathrm{Conv2D}^{PW}_1(X), \\
    A &= \mathrm{Conv2D}^{DW}(\mathrm{GELU}(\mathrm{Conv2D}^{PW}_2(X))) \\
    Z &= \mathrm{Conv2D}^{PW}_3(A \otimes V)
\end{align}

where $\otimes$ stands for the element-wise product, while $\mathrm{Conv2D}^{PW}_{[1:3]}$ represent three different point-wise convolution layers. Additionally, $\mathrm{Conv2D}^{DW}$ signifies a depth-wise convolution with an $11 \times 11$ kernel size.

Convolutional attention allows each spectral feature to establish correlations with all others in the $11 \times 11$ window and the interaction among spectral channels is enabled by channel projection. Following the Transformer, the remainder of the Convolutional Modulation uses skip connections, channel normalization as the norm, and point-wise convolution as the feedforward. In comparison to self-attention, Convolutional Modulation takes advantage of convolutions, leading to improved memory overhead for long audio processing. 

Then local attention with a window size of three is implemented to focus on extracting sub-band information. This is achieved by conducting dot-product attention on three neighboring frequency bands for each given band, allowing the model to capture local-level context.
Following this step, TCN is utilized to reconstruct the time-domain information, ensuring a comprehensive representation of the original signal's temporal dynamics.

\subsection{Mask/Bias Decoder} 
            
As shown in Fig.~\ref{fig:encoder} (b), the Mask/Bias Decoder begins with a depth-separable convolution, followed by a dual-path convolutional structure. Each path comprises a unique point-wise convolution; one is activated by a sigmoid function, and the other is activated by a $tanh$ function. Subsequently, the outputs of both paths are combined through an element-wise product, followed by another point-wise convolution, LN, and sigmoid activation.

\subsection{Loss Function}

The loss function takes into account both the spectrogram and the magnitude spectrum of the speech signal, which can be represented as follows:

\begin{gather}
    \mathcal{L}^{Mag}=\left \| M^* -  M   \right \|^2,\\
    \mathcal{L}^{RI}=\left \|S^*_r - \widetilde{S}_r \right \|^2 +\left \|S^*_i - \widetilde{S}_i \right \|^2,\\
    \mathcal{L}=0.5 \mathcal{L}^{Mag}+0.5 \mathcal{L}^{RI} 
\end{gather}

Above, loss function $\mathcal{L}$ can be represented as a combination of two individual loss terms, one for the magnitude and the other for spectrogram. The first term $\mathcal{L}^{Mag}$ measures the squared error between the target magnitude spectrum $M^*$ and the estimated magnitude spectrum $M$. The second term $\mathcal{L}^{RI}$ measures the squared error between the target and estimated spectrogram, namely the real components $S^*_r$ and $\widetilde{S}_r$, and the imaginary components $S^*_i$ and $\widetilde{S}_i$. The overall loss function $\mathcal{L}$ combines the individual loss terms $\mathcal{L}^{Mag}$ and $\mathcal{L}^{RI}$ with equal weights, ensuring that both the magnitude and spectrogram contribute to the total error.

\section{EXPERIMENTS}

\subsection{Data}

To evaluate our model, we use the publicly available Voice Bank + DEMAND dataset, which is a standard dataset for speech enhancement evaluation.
The Voice Bank + DEMAND dataset, as proposed by \cite{valentini2016investigating}, combines clean speech recordings from the Voice Bank corpus \cite{veaux2013voice} with various acoustic noise recordings from the Diverse Environments Multichannel Acoustic Noise Database (DEMAND) \cite{thiemann2013diverse} dataset, each recorded with 48kHz sampling rate.
The training set consists of 11,572 noisy-clean pairs, created by mixing clean speech from 28 Voice Bank speakers with 10 types of noise (2 synthetic and 8 from DEMAND) at four different SNR settings $\left\{0 \ \rm{dB}, 5 \ \rm{dB}, 10 \ \rm{dB}, 15 \ \rm{dB}\right\}$, these settings result in 40 different patterns.
For evaluation, the test set contains 824 noisy-clean pairs which were created with 2 unseen speakers and 5 unseen noise types at different SNR settings $\left\{2.5 \ \rm{dB}, 7.5 \ \rm{dB}, 12.5 \ \rm{dB}, 17.5 \ \rm{dB}\right\}$.
This arrangement helps to create a more challenging evaluation setup, as the model are exposed to diverse acoustic conditions and speaker variations.

\subsection{Implementation setup}

All audio samples are resampled from 48kHz to 16kHz and trimmed to a maximum length of 3 seconds. By applying a 320-point STFT with a 25 ms Hanning window length and a 10 ms hop length, we obtain spectrograms containing 161-dimensional spectral characteristics. Following the power compression technique presented in \cite{li2021importance}, we apply it to the magnitude using a coefficient of $0.5$, while leave the phase unchanged. We adopt the Adam optimiser \cite{kingma2014adam} with a learning rate of 5e-4, $\beta_1$ of 0.95, and $\beta_2$ of 0.999, train the model for 50 epochs with a batch size of 4.

\subsection{Evaluation Metrics}

To evaluate our model, we use several evaluation metrics as follows:

\begin{itemize}
\item \textbf{WB-PESQ \cite{union2007wideband} (Wideband Perceptual Evaluation of Speech Quality):} WB-PESQ is an extension of the PESQ \cite{rix2001perceptual} metric for wideband speech signals. It is a widely used objective measure that predicts the subjective quality of speech signals. It ranges from -0.5 to 4.5, with higher scores indicating better perceived quality.

\item \textbf{STOI \cite{taal2011algorithm} (Short-Time Objective Intelligibility):} STOI is an objective metric that gauges speech intelligibility. The score ranges from 0 to 100\%, with higher percentages denoting better intelligibility.

\item \textbf{CSIG \cite{hu2007evaluation} (Composite Signal-to-Noise Ratio):} CSIG evaluates the overall quality of the enhanced signal, taking into account both distortion and noise suppression. Higher scores indicate better signal quality.

\item \textbf{CBAK \cite{hu2007evaluation} (Composite Background-to-Noise Ratio):} CBAK measures the effectiveness of background noise suppression. Higher scores signify more effective noise suppression.

\item \textbf{COVL \cite{hu2007evaluation} (Composite Overall Quality):} COVL is a comprehensive metric that considers both signal quality and noise suppression. Higher scores indicate better overall performance.

\item \textbf{SSNR (Segmental Signal-to-Noise Ratio):} SSNR is an objective metric that assesses the quality of the enhanced speech signal by comparing it to the original clean signal. Higher SSNR scores indicate better speech enhancement.

\end{itemize}

\subsection{Comparison With Previous Methods}

\begin{table*}[tbh]
  \caption{This table compares the performance of our proposed model with previous methods on the Voicebank+Demand test set. Missing values are denoted by \textbf{-}.}
  \vspace{.2em}
  \begin{small}
  \label{tab:voicebank_test}
  \robustify\bfseries
  \sisetup{
    table-number-alignment = center,
    table-figures-integer  = 1,
    table-figures-decimal  = 2,
    table-auto-round = true,
    detect-weight = true
  }
\begin{minipage}[c]{\textwidth}
\centering
\begin{tabular}{rl | S | S | S | S S S | S}
      \toprule%
      & \textbf{Model}  & \textbf{Params(M)} & \textbf{WB-PESQ} & \textbf{STOI(\%)} & \textbf{CSIG} & \textbf{CBAK} & \textbf{COVL} & \textbf{SSNR} \\%
      \cmidrule{2-9}%
      \cmidrule{2-9} \multicolumn{9}{c}{\textbf{Unprocessed}} \\ \cmidrule{2-9}
      & Noisy & \text{-} & 1.97 &  92.1 & 3.35 & 2.44 & 2.63 & 1.68 \\%
      \cmidrule{2-9} \multicolumn{9}{c}{\textbf{Conventional Statistical-based Baseline}} \\ \cmidrule{2-9}
      & Wiener \cite{scalart1996speech} & \text{-} & 2.22 & \text{-} & 3.23 & 2.68 & 2.67 & 5.07 \\
      \cmidrule{2-9} \multicolumn{9}{c}{\textbf{Previous Deep Learning Models}} \\ \cmidrule{2-9}
      & RNNoise \cite{valin2018hybrid} & \textbf{0.06} & 2.34 & 92.2 & 3.40 &  2.51 & 2.84  & \text{-} \\%
      & Wavenet \cite{rethage2018wavenet} & \textbf{-} & \textbf{-} & \textbf{-} & 3.62 & 3.23 & 2.98 & \textbf{-} \\%
      & SETransformer \cite{yu2022setransformer} & \text{-} & 2.62 & 93.0 & \text{-} & \text{-} & \text{-} & \text{-} \\%
      & NSNet2 \cite{braun2021towards} & 6.17 & 2.47 & 90.3 & 3.23 & 2.99 & 2.90 & \text{-} \\%
      & PercepNet \cite{valin2020perceptually} & 8.00 & 2.73 & \text{-} & \text{-} & \text{-} & \text{-} & \text{-} \\%
      & DCCRN \cite{hu2020dccrn} & 3.70 & 2.54 & 93.8 & 3.74 & 3.13 & 2.75 & \text{-} \\%
      & DCCRN+ \cite{lv2021dccrn+} & 3.30 & 2.84 & \text{-} & \text{-} & \text{-} & \text{-} & \text{-} \\%
      & S-DCCRN \cite{lv2022s} & 2.34 & 2.84 & 94.0 & 4.03 & \textbf{3.43} &  2.97 & \text{-} \\%
      & FullSubNet+ \cite{chen2022fullsubnet+} & 8.67 & \textbf{2.88} & 94.0 & 3.86 & 3.42 & \textbf{3.57}  & \text{-} \\%
      & DeepFilterNet \cite{schroter2022deepfilternet} & 1.78 & 2.81 & \textbf{94.2} & \textbf{4.14} & 3.31 & 3.46 & \text{-} \\%
      \cmidrule{2-9}
      \cmidrule{2-9} \multicolumn{9}{c}{\textbf{Ours}} \\ \cmidrule{2-9}
      & Spectrum Attention Fusion & \textbf{0.58} & 2.84 & 94.3 & 4.04 & 3.38 & 3.43 & 9.05 \\%
      & \qquad- Phase Spectrum Input & \textbf{0.58} & 2.71 & 94.1 & 3.87 & 3.29 & 3.28 & 8.80 \\%
      & \qquad+ 1 Attention Fusion Layer & 1.16 & 2.78 & 94.1 & 3.91 & 3.31 & 3.35 & 8.92 \\%
      & \qquad\qquad+ skip connection & 1.16 & \ \textbf{2.89} & \textbf{94.4} & \textbf{4.18} & \textbf{3.43} & \textbf{3.59} & \textbf{9.10} \\%
      \cmidrule{2-9}%
    \end{tabular}
  \end{minipage}
  \end{small}
\end{table*}%

As reported in Table \ref{tab:voicebank_test}, we provide a comprehensive comparison of our proposed model with previous methods on the Voicebank+Demand test set.
Our model exhibits competitive performance when compared to several previous deep learning models. 
Specifically, the Spectrum Attention Fusion achieves a WB-PESQ score of 2.84, outperforming the SETransformer model that employs a vanilla Transformer. This score is also on par with leading models such as FullSubNet+ and S-DCCRN. Moreover, our model attains the highest STOI score of 94.3\% among all evaluated methods. 
It is important to note that our model achieves these high performance scores with only 0.58 million parameters, which is significantly lower than many previous models. 
Furthermore, the ablation studies in Table \ref{tab:voicebank_test} demonstrate the impact of different components in our model. For instance, integrating the phase spectrum input contributes to an improvement in the model's performance. Interestingly, merely increasing the number of attention fusion layers does not necessarily lead to improved performance. On the other hand, combining the attention fusion layer with skip connections results in a boost in performance.

\subsection{The Influence of Phase Spectrum Input}

\begin{figure}[h]
    \vspace{-0.55cm}
    \centering
    \centerline{\includegraphics[width=0.87\columnwidth]{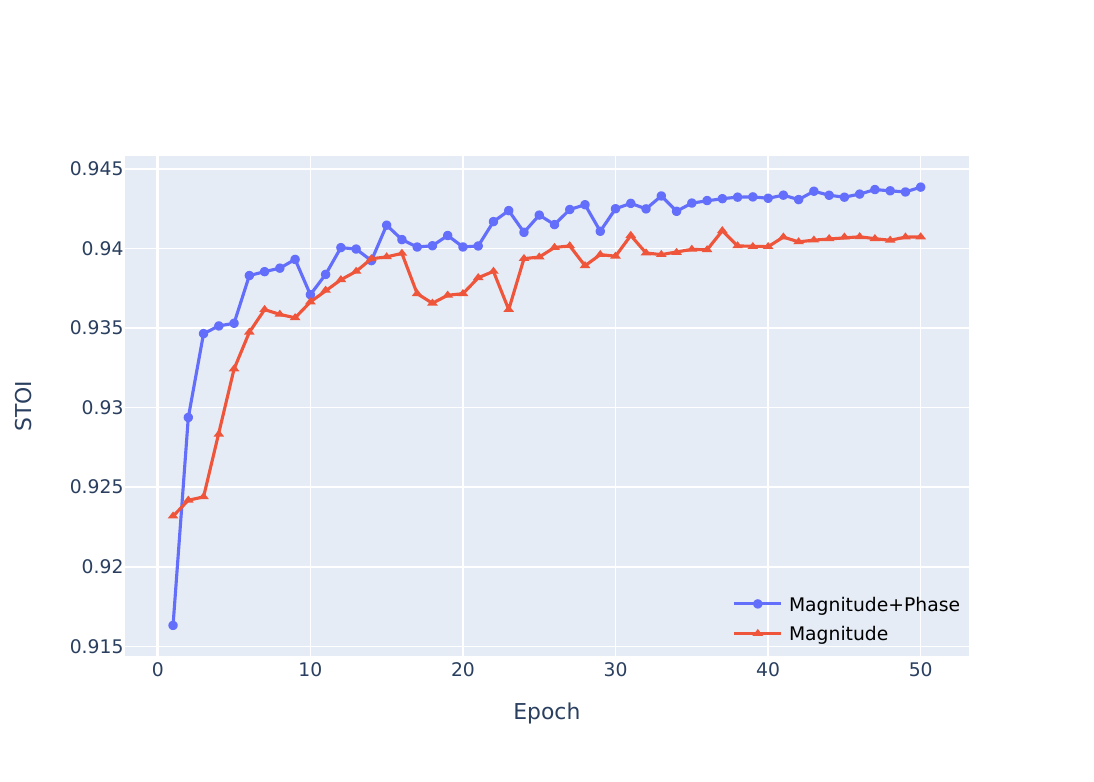}}
    \vspace{-0.6cm}
    \caption{
        \label{fig:mag_vs_mag+pha}
        Comparison of STOI performance during training over 50 epochs for models using Magnitude-only and Magnitude+Phase spectrum inputs. 
        }
\end{figure}

It is important to note some previous work such as \cite{li2021two} only used the magnitude spectrum as input.
To see how much improvement do phase spectrum provide, we also train our models with different combinations inputs: only magnitude and magnitude combined with phase. We then compared their STOI performance during 50 epochs of training. Fig.~\ref{fig:mag_vs_mag+pha} demonstrate that the model with both magnitude and phase inputs outperforms the magnitude-only model in terms of STOI.

\subsection{Scalability}

As shown in Table \ref{tab:voicebank_test}, we also explore the scalability of our proposed model by examining the effects of increasing the number of Attention Fusion layers and incorporating skip connections in the architecture. 
We found that merely increasing the number of layers in the Attention Fusion mechanism led to decreased performance. This outcome suggests that simply deepening the model does not ensure improved results, and can even make the learning process more challenging. The drop in performance might be due to the vanishing gradient problem, which causes gradients to become too small for effective weight updates during back-propagation. On the other hand, when we employed skip connections while increasing the number of layers, the model's performance improved. This finding highlights the benefits of using skip connections in deeper architectures to mitigate the vanishing gradient problem and facilitate gradient flow through the network. 
These results emphasise the need for careful architectural design and optimisation to balance complexity and performance. Furthermore, the results demonstrate the scalability of our Spectrum Attention Fusion model, which can be improved further by leveraging skip connections and other architectural innovations.

\section{Conclusion}

In this work, we proposed an efficient model named Spectrum Attention Fusion for monaural speech enhancement. 
With a compact design, our model achieves competitive performance with fewer parameters than previous approaches.
The incorporation of phase spectrum input and key components such as Attention Fusion proved to be instrumental in enhancing the effectiveness of our model.
By providing high-quality speech enhancement with an efficient architecture, our work contributes to the development of more streamlined models for speech enhancement applications.

%
%
%
\bibliographystyle{splncs04}
\bibliography{reference}
%





\end{document}